\documentclass[11pt,twoside]{article}
\usepackage{FUSE2004}
\usepackage{natbib}

\usepackage{epsf}
\usepackage{psfig}
\usepackage{lscape}
\begin{document}

\title{A Lyman Continuum Explorer - LyContEx}
\author{Stephan R. McCandliss}
\affil{The Johns Hopkins University }
\begin{abstract}
  
{ \em GALEX} 
has identified large numbers of extra galactic objects that are too
faint to be observed by {\em FUSE}.  These star forming systems are
potential sources of Lyman continuum (Lyc) radiation and may contribute to
the metagalactic background radiation that ionizes most of the baryons
in the universe.  By measuring the level of Lyc radiation
leaking from these galaxies we can quantify the contribution of stars,
relative to quasars, to the ionizing background at low redshift ($0.025
\la z \la 0.4$) and provide insight into the reionization processes at
high redshift.  A meter class instrument with a moderate spectral
dispersion, wide-field,
multi-object spectrograph can detect Lyc leak from a
statistically significant number of these galaxies, if the
average escape fraction $ f_{esc} \ga$ 1\%.  Such an instrument will
have enormous discovery potential because of the large number of
extra-galactic objects that exist within its grasp.  LyContEx  will be
scientifically productive, synergistic with existing surveys, and can be
developed into a far ultraviolet mission in the near future.

\end{abstract}

\vspace*{-.5in}
\section{ Introduction }
\vspace*{-.1in}

For all its considerable success {\em FUSE} has three limitations.  The
spectral resolution is too low, the spectral resolution is too high and
the \'{e}tendue ($A_{eff}\Omega$, grasp, information content)  is limited to
a single object.

The challenge for future far ultraviolet missions is to identify
compelling new science opportunities enabled by instrumentation that
overcomes these limitations.  Advances in high resolution spectroscopy
($R \ge$ 20000) will require a very large telescope  \citep{Moos:2004}, a
new approach to spectrograph design, and a multi-decadal timescale for
development.  A meter class, high \'{e}tendue, moderate dispersion far
ultraviolet spectrograph will have a shorter development timescale and
an enormous potential for discovery because of the large number of
extra galactic sources within its grasp too faint to be observed by
{\em FUSE}.  One new science opportunity enabled by such an instrument
will be the search for the elusive Lyc radiation leaking from low
redshift star-forming galaxies.

\citet{McCandliss:2004} have detailed a new moderate dispersion, wide
field, far ultraviolet spectroscopic survey instrument incorporating an
array of micro-shutters \citep{Moseley:2000} designed to reach
significantly lower sensitivities than {\em FUSE} and increase the
information gathering power.  Observations of objects with flux levels
below the {\em FUSE} background equivalent flux (BEF $\sim 5 \times
10^{-15} \:{\rm ergs\:cm^{-2} s^{-1}\:\AA^{-1}}$),  face an exposure
time ``Wall'' requiring increasingly deep exposures.  A LyContEx
mission with  $R \sim$ 3000  will yield a BEF $\sim 2 \times 10^{-17}
\:{\rm ergs\:cm^{-2}\:s^{-1}\:\AA^{-1}}$.  \citet{Xu:2004} finds  $>$
850 gal $(\deg)^{-2}$ brighter than  23 mag$_{AB}$ = $3 \times 10^{-17}
\:{\rm ergs\:cm^{-2}\:s^{-1}\:\AA^{-1}}$ at 1540 \AA\ from {\em GALEX}
number counts (Figure 1).  I show here a statistically significant number of
these galaxies will have detectable levels of Lyc if the $ f_{esc} \ga$
1\%.

\begin{figure*}
\vspace*{.2in}
\psfig{file=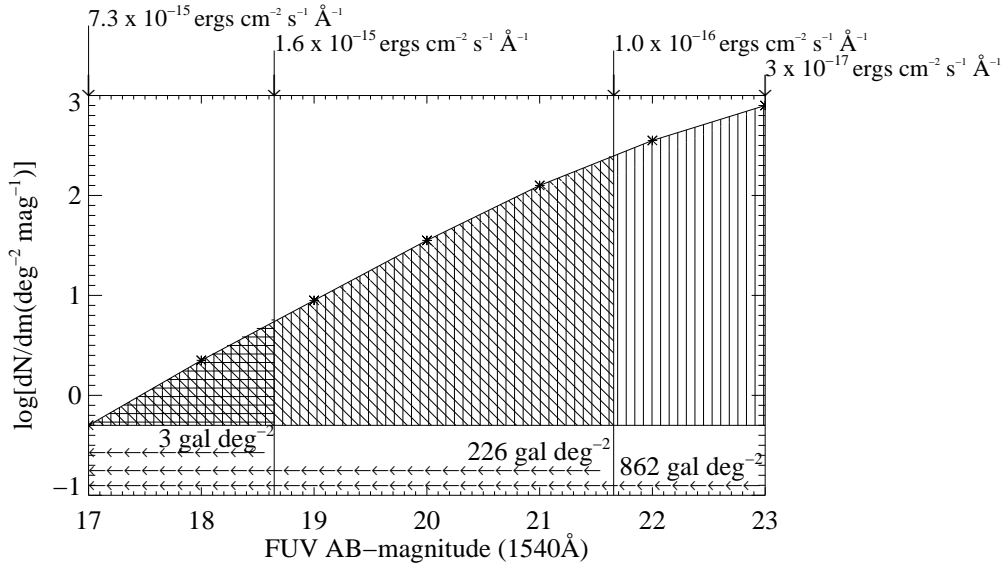,angle=90,width=4.5in}
\caption[]{Differential number counts  
from {\em GALEX} observations. The cross
hatched areas have been integrated to determine the total number of
galaxies brighter than the limiting fluxes on the top of the
figure. The integrated number of galaxies per $(\deg)^{2}$ are
printed under the cross-hatchings. }
\label{numcts}
\vspace*{-.2in}
\end{figure*}

\section{Why Seek Lyc Escaping from Galaxies?}
\vspace*{-.1in}

The escape fraction is a key parameter widely used to ``fine tune'' the
ionization history of the universe \citep{Madau:1996}. It is widely
thought that the ionizing background that permeates metagalactic space
is produced by quasars and star-forming galaxies
\citep{Madau:1996,Shull:1999,Heckman:2001}, yet the contribution from
star-forming galaxies has yet to be directly and convincingly
detected.  At low redshift upper limits to the escape fraction range
from 5 - 50\% \citep{Leitherer:1995,Hurwitz:1997,Heckman:2001,Deharveng:2001}. At high redshift \citet{Steidel:2001} have claimed a tentative detection
in $z$ = 3.4 Lyman break galaxy composite spectrum.

A quantitative assessment of the contribution of galaxies to the
ionizing background will require a large number of observations before
a Lyc luminosity function can be established \citep{Deharveng:1997}.
\citet{Heckman:2001} point out that determining the sources,
strength and evolution of the metagalactic radiation field is crucial
to understanding the fundamental properties of the IGM where the bulk
of the baryons in the universe reside.  Association of  individual
escape fractions within the mix of galactic morphological/spectral
types found in the early universe, would allow the ionizing background
at high redshift to be calculated semi-empirically at an epoch where
starbursts are expected to dominate the reionization process.  This
would constrain the allowable fine tuning in ionization history
models.

\begin{figure*}
\psfig{file=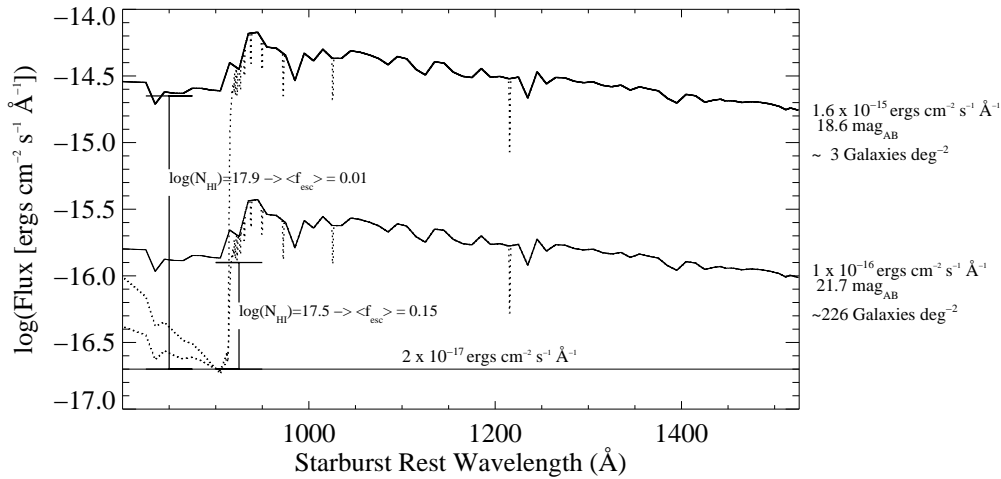,angle=90,width=4.5in}
\vspace*{-.2in}
\caption[]{A Starburst99 model (solar metalicity, Salpeter IMF, upper mass cutoff of
100 $M_{ \sun }$) is used to estimate the minimum
observable {\em GALEX} FUV flux for a Lyc flux
at the background limit of the proposed LyContEx for average escape
fractions  1 and 15\%.}
\label{fortis}
\vspace*{-.2in}
\end{figure*}

\section{Predicting the Number of Detectable Lyc Emitters }
\vspace*{-.1in}

{\em GALEX} FUV differential number counts of galaxies and Starburst99
models \citep{Leitherer:1999} of continuous star formation  can be used
to estimate the number of detectable galaxies emitting Lyc photons.  
An escape fraction of 1 and 15\% is imposed by applying the appropriate
neutral hydrogen absorption model to a Starburst99  
continuous star formation model. Offsetting the resulting spectra at
$F_{900}$ to the expected LyContEx BEF, the limiting {\em
GALEX} FUV flux can be found.
Integration of the differential number counts to these limiting
$F_{1540}$ fluxes provides the prediction of the total number of
star-forming galaxies that should be detectable (see Figure 1).  I find for the whole sky $\sim$ 10$^{5}$ detectable galaxies if $f_{esc}$ = 1\% and $\sim$
10$^{7}$ galaxies if $f_{esc}$ = 15\% .  These numbers
should be sufficient to allow the determination of a Lyc luminosity function.


\acknowledgements I wish to thank Erik Wilkinson, Karl Glazebrook, Ivan
Baldry, Jeffrey Kruk and Warren Moos, Luciana Bianchi,
Gerhardt Meurer, Bill Blair, Paul Feldman and Kevin France for useful
discussions and
their support.  This work is supported through NASA grant NNG04WC03G
to JHU.

\vspace*{-.2in}

\bibliography{references04}
\bibliographystyle{apj}


\end{document}